\documentclass{Interspeech}

\interspeechcameraready

\title{TS-URGENet: A Three-stage Universal Robust and Generalizable Speech Enhancement Network}

\author[affiliation={1,2}]{Xiaobin}{Rong}
\author[affiliation={1,2}]{Dahan}{Wang}
\author[affiliation={1,2}]{Qinwen}{Hu}
\author[affiliation={1,2}]{Yushi}{Wang}
\author[affiliation={2}]{Yuxiang}{Hu}
\author[affiliation={1,2}]{Jing}{Lu}

\affiliation{Key Laboratory of Modern Acoustics}{Nanjing University}{China}
\affiliation{NJU-Horizon Intelligent Audio Lab}{Horizon Robotics}{China}
\email{\{xiaobin.rong, dahan.wang, qinwen.hu, yushi.wang\}@smail.nju.edu.cn, yuxiang.hu@horizon.cc, lujing@nju.edu.cn}

\usepackage{comment, amsmath, hyperref}
\usepackage{makecell}
\usepackage{tabularx}
\usepackage{array}

\begin{document}

\maketitle

\begin{abstract}
    Universal speech enhancement aims to handle input speech with different distortions and input formats. To tackle this challenge, we present TS-URGENet, a \textbf{T}hree-\textbf{S}tage \textbf{U}niversal, \textbf{R}obust, and \textbf{G}eneralizable speech \textbf{E}nhancement \textbf{Net}work. To address various distortions, the proposed system employs a novel three-stage architecture consisting of a filling stage, a separation stage, and a restoration stage. The filling stage mitigates packet loss by preliminarily filling lost regions under noise interference, ensuring signal continuity. The separation stage suppresses noise, reverberation, and clipping distortion to improve speech clarity. Finally, the restoration stage compensates for bandwidth limitation, codec artifacts, and residual packet loss distortion, refining the overall speech quality.
    Our proposed TS-URGENet achieves outstanding performance in the Interspeech 2025 URGENT Challenge, ranking 2nd in Track 1.
    
\end{abstract}

\section{Introduction}

Conventional speech enhancement (SE) primarily focuses on eliminating undesired signals such as noise and reverberation. In contrast, a more general SE task proposed in the Speech Signal Improvement (SSI) Challenges aims to improve speech quality by considering a broader range of factors, including coloration, discontinuity, and loudness, in addition to noise and reverberation \cite{SIG1, SIG2}. Recently, the concept of universal speech enhancement (USE) has been proposed. Compared to general SE, USE considers using a single model to handle diverse input formats, including variable channels, lengths, and sampling frequencies, in addition to various distortions \cite{USE}. In the Interspeech 2025 URGENT Challenge\footnote{\url{https://urgent-challenge.github.io/urgent2025/}}, a concrete task for USE is proposed: to build a single model capable of addressing seven types of distortions while accommodating sampling frequency flexibility. The distortions include additive noise, reverberation, clipping, bandwidth limitation, codec artifacts, packet loss, and wind noise. The supported sampling frequencies include 8, 16, 22.05, 24, 32, 44.1, and \SI{48}{kHz}.

To build a USE system flexible with different sampling frequencies, one straightforward solution is to always upsample the speech to \SI{48}{kHz} as pre-processing, allowing the model to process data with a fixed sampling frequency. The output is then downsampled to the original sampling frequency as the final enhanced speech. Alternatively, sampling-frequency-independent (SFI) approaches can directly adapt to various sampling frequencies \cite{Toward_USE}. 
Extensive research has been conducted on addressing individual distortion types separately \cite{Declipping, AP-BWE, Codec_enhancement, BS-PLCNet}. However, recent advances have revealed that various distortion types can be effectively handled using a unified framework, which can be classified into two categories: single-stage and multi-stage methods. In single-stage approaches, all distortions are addressed simultaneously. For instance, VoiceFixer \cite{Voicefixer} employs ResUNet \cite{ResUNet} to address additive noise, reverberation, clipping, and bandwidth limitation in the Mel domain, and utilizes a neural vocoder to synthesize a waveform. In \cite{USE_diffusion}, a score-based diffusion model is proposed to tackle 55 different distortions at the same time. In multi-stage approaches, distortions are grouped into different stages based on their characteristics. In \cite{Clipping_codec_gaps}, a pipeline cascades a time-domain U-Net for clipping and packet loss with a time-frequency U-Net for codec distortions. In the winning solutions of the SSI Challenges, two-stage frameworks are widely employed, where the first stage addresses various distortions simultaneously, and the second stage eliminates residual noise and artifacts produced by the first stage \cite{SIG1_Gesper, SIG2_KSNet, SIG2_RADNet, SIG2_NJUNet}.

The complex acoustic scenarios in the URGENT Challenge, where speech can be contaminated by strong noise, reverberation, and other distortions, make a single-stage framework highly challenging. This limitation can hinder the model's generalization ability and ultimately degrade enhancement performance. On the other hand, the two-stage ``restoration and enhancement" framework performs more effectively in high signal-to-noise ratio (SNR) scenarios. When applied to data with strong noise and reverberation, it may inadvertently generate noise instead of restoring speech, resulting in additional artifacts in the processed output. Considering the distinct characteristics of the seven targeted distortions, we propose a three-stage universal, robust, and generalizable speech enhancement network named TS-URGENet, which divides the complex enhancement into three stages: filling, separation, and restoration. Specifically, the filling stage fills the gaps caused by packet loss distortion; the separation stage simultaneously handles noise, reverberation, and clipping distortion; and the restoration stage addresses bandwidth limitation, codec artifacts, and residual packet loss distortion. Our contributions are as follows:
\begin{itemize}
    \item We propose a novel filling-separation-restoration framework for universal speech enhancement, to address various distortions with different input sampling frequencies.
    \item We introduce an effective metric-aware fine-tuning loss function to directly and collaboratively optimize multiple target metrics.\looseness=-1
    \item Our proposed TS-URGENet achieves outstanding performance in the Interspeech 2025 URGENT Challenge, ranking 2nd in Track 1.
\end{itemize}

\section{Method}
\begin{figure}[t]
    \centering
    \includegraphics[width=\linewidth]{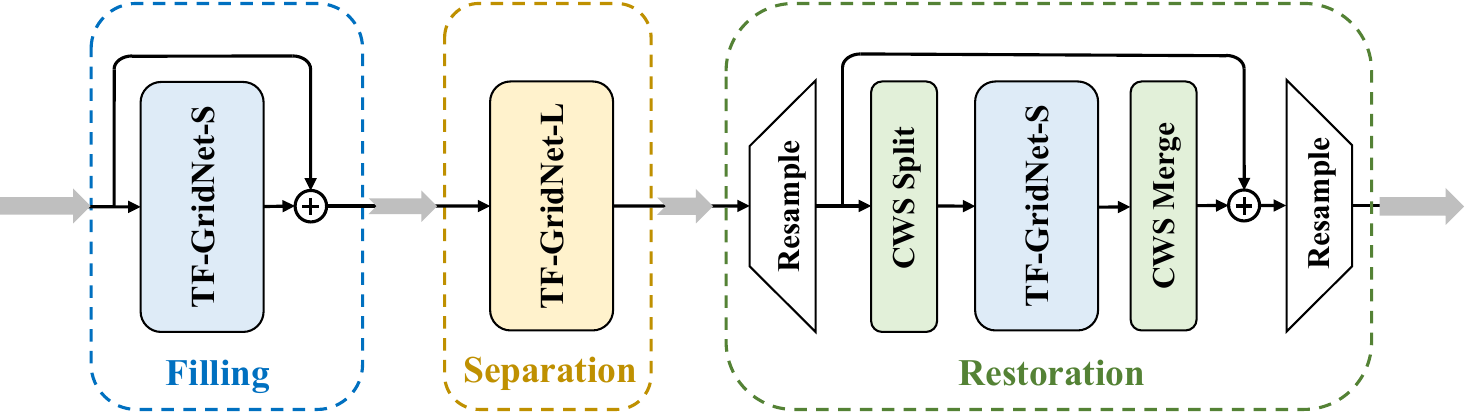}
    \caption{Overview of the TS-URGENet framework.}
    \label{fig:ts-urgenet}
\end{figure}

As shown in Fig. \ref{fig:ts-urgenet}, our proposed framework is composed of three parts: filling, separation, and restoration stages. We first employ a small-scale TF-GridNet \cite{TF-GridNet} (TF-GridNet-S) as the filling module to fill the gaps caused by packet loss distortion. A residual connection is added after the filling stage to minimize information loss. Subsequently, a larger-scale TF-GridNet (TF-GridNet-L) is applied as the separation module for denoising, dereverberation, and declipping. Finally, we sequentially integrate the channel-wise subband (CWS) \cite{CWS} splitting operation, TF-GridNet, and the CWS merging operation to create CWS-TF-GridNet as the restoration module, which performs bandwidth extension (BWE), codec artifact reduction, and further packet loss concealment (PLC). 

As TF-GridNet inherently satisfies the sampling-frequency-independent (SFI) requirement, both the filling and separation stages are adaptive to various sampling frequencies. However, due to the constraints imposed by the CWS operation, CWS-TF-GridNet is restricted to processing signals at a fixed sampling frequency. To address this limitation, resampling modules are incorporated at both the input and output of the restoration stage. All the stages operate at the time-frequency (T-F) domain, the short-time Fourier transform (STFT) and inverse STFT (iSTFT) are integrated at the input and output of the network in each stage, while omitted in Fig. \ref{fig:ts-urgenet} for simplicity. 

\subsection{Filling stage}
\label{sec:filling}
We found that speech discontinuity caused by packet loss can degrade the performance of the separation module. Therefore, we first incorporate a filling stage, which is expected to restore the components of the lost regions while maintaining the integrity of the other regions. Although in practice, we found that the model often fills the gaps with noise rather than valid speech, this approach helps improve the continuity of signals, which in turn facilitates the subsequent separation stage.

We use a GAN-based approach to perform the filling. TF-GridNet-S is employed as the generator.  A residual connection is added afterward to minimize information loss. As for the discriminators, the multi-period discriminator (MPD) and the multi-resolution discriminator (MRD) used in BigVGAN \cite{BigVGAN} are adopted together. We employ the same training loss as in Gesper \cite{SIG1_Gesper}, consisting of reconstruction, adversarial, and feature-matching terms, with respective weights of 20, 1, and 1.

\subsection{Separation stage}
Inspired by \cite{Declipping}, we group clipping, noise, reverberation, and wind noise together to be handled during the separation stage. TF-GridNet-L is employed to achieve high-performance separation.

We adopt an effective metric-aware fine-tuning (MAFT) paradigm. In the first step, the model is trained using a conventional loss function applied to both waveform and spectrogram domains, which consists of a signal-to-distortion ratio (SDR) term, a log-spectral distance (LSD) term, and multiple mean square error (MSE) terms:
\begin{equation}
  \mathcal{L}_1=2\mathcal{L}_{\text{SDR}}+ 1.5\mathcal{L}_{\text{LSD}} + 70 \mathcal{L}_{\text{mag}} + 30\left(\mathcal{L}_{\text{real}} + \mathcal{L}_{\text{imag}} \right)
\end{equation}
with each term calculated as:
\begin{gather}
    \mathcal{L}_{\text{SDR}} = -\log_{10} (\frac{||s||^2}{||s-\hat{s}||^2})  \\
    \mathcal{L}_{\text{LSD}} = \frac{1}{T} \sum_{t=1}^{T} \sqrt{ \frac{1}{F} \sum_{f=1}^{F} \left( \log_{10} ( \frac{|S|}{|\hat{S}|} ) \right)^2 } \\
    \mathcal{L}_{\text{mag}} = \mathrm{MSE}(|S|^{0.3}, |\hat{S}|^{0.3})  \\
    \mathcal{L}_{\text{real}} = \mathrm{MSE} (S_r/|S|^{0.7}, \hat{S_r}/|\hat{S}|^{0.7})  \\
    \mathcal{L}_{\text{imag}} = \mathrm{MSE} (S_i/|S|^{0.7}, \hat{S_i}/|\hat{S}|^{0.7}) 
\end{gather}
where $T$ and $F$ denote the time and frequency dimensions of the spectrogram, respectively. $S$ and $\hat{S}$ are the clean and enhanced spectrograms, respectively. $S_r$, $S_i$, $\hat{S_r}$, and $\hat{S_i}$ represent the real and imaginary parts of the clean and enhanced spectrograms, respectively. 

After convergence, the model is fine-tuned for a few additional epochs using our elaborately designed metric-aware loss functions. We implement various metric-aware loss functions, including Mel cepstral distortion (MCD) \cite{MCD}, perceptual evaluation of speech quality (PESQ) \cite{PESQ}, DNSMOS \cite{DNSMOS-P835}, and UTMOS \cite{UTMOS}. The MCD-aware term is computed as the MSE between the Mel-frequency cepstral coefficients (MFCC) of clean and enhanced speech. For the PESQ-aware term, we adopt a differentiable implementation\footnote{\url{https://github.com/audiolabs/torch-pesq}}. As for the DNSMOS and UTMOS-aware terms, we utilize corresponding pre-trained models from the official evaluation scripts\footnote{\url{https://github.com/urgent-challenge/urgent2025_challenge/blob/main/evaluation_metrics}} for loss calculation.

For downstream-task-related metrics including phoneme similarity (PhnSim) \cite{PhnSim}, SpeechBERTScore \cite{SpeechBERTScore}, speaker similarity (SpkSim), and character accuracy (CAcc), we use the representations of WavLM \cite{WavLm} to construct a loss function, aiming to fit these metrics. We use the continuous distillation loss function proposed in \cite{SpeechTokenizer} to measure the similarity between clean and enhanced speech representations, formulated as:
\begin{equation}
    \mathcal{L}_{\text{WavLM}}=-\frac{1}{T}\frac{1}{D}\sum_{t=1}^{T}\sum_{d=1}^{D}\log_{10}(\sigma(\mathrm{cos}(R, \hat{R})))
\end{equation}
where $R$ and $\hat{R}$ denote the WavLM representations of clean and enhanced speech, respectively, $D$ represents the feature dimension, $\sigma(\cdot)$ denotes the sigmoid activation, and $\cos(\cdot,\cdot)$ represents cosine similarity. We utilize the WavLM Base+ model\footnote{\url{https://github.com/microsoft/unilm/tree/master/wavlm}} for feature extraction.

The complete metric-aware loss function is formulated as a weighted combination of $\mathcal{L}_1$ and the aforementioned components:\looseness=-1
\begin{align}
    \mathcal{L}_2 = &\mathcal{L}_1 + 0.004\mathcal{L}_{\text{MCD}} + 0.5\mathcal{L}_{\text{PESQ}} \\
    + &0.5\mathcal{L}_{\text{UTMOS}} + 0.4\mathcal{L}_{\text{DNSMOS}} + 2.5\mathcal{L}_{\text{WavLM}}
\end{align}
We conduct an exhaustive investigation to determine the components that are retained in the final loss function, which will be discussed in Sec. \ref{sec:maft}.

\subsection{Restoration stage}
In this stage, distorted speech is expected to recover from bandwidth limitations, codec artifacts, and residual packet loss. The speech is first resampled to \SI{48}{kHz} and then processed by our proposed CWS-TF-GridNet. After applying the STFT, the fullband complex spectrogram is divided into three subbands, which are concatenated along the channel dimension as input. The merging procedure mirrors the splitting process: the network output is split channel-wise into predictions for each subband, and then concatenated along the frequency dimension. The output undergoes the iSTFT and is then resampled back to its original sampling frequency. A skip connection is also added at both ends of the stage to minimize information loss. GAN is also employed in the restoration stage, where CWS-TF-GridNet serves as a generator while MRD and multi-band discriminators (MBD) proposed in \cite{SIG1_Gesper} are utilized to improve the perceptual quality of the generated speech. The loss function in this stage is consistent with that described in Sec. \ref{sec:filling}.

In a multi-stage framework training, a fine-tuning strategy is commonly adopted to accelerate convergence \cite{SIG2_RADNet}. Initially, the model is trained independently on distorted data, which is simulated by corrupting clean speech with specific distortions. Upon reaching convergence, the model is further fine-tuned using the output from the previous stage to mitigate domain mismatch. This approach is termed stage-aware fine-tuning (SAFT) in this paper. In practice, we observe that while SAFT enhances overall speech quality to some extent, it significantly degrades certain intrusive metrics. To circumvent this drawback, we propose a novel approach that integrates SAFT with MAFT to form a joint stage and metric-aware fine-tuning (JFT) strategy. Specifically, JFT utilizes data output from previous models for training, similar to SAFT; however, it adopts the metric-aware loss employed in MAFT, while incorporating additional adversarial and feature-matching components:
\begin{equation}
    \mathcal{L}_3 = 10\mathcal{L}_2 + \mathcal{L}_{\text{adv}} + 0.2\mathcal{L}_{\text{feat}}
\end{equation}

\section{Experiments}
\subsection{Datasets}
We use the data provided in Track 1 of the Interspeech 2025 URGENT Challenge and employ the official scripts\footnote{\url{https://github.com/urgent-challenge/urgent2025_challenge/tree/main/simulation}} for simulation. The training data are dynamically simulated during the training process. The specific configuration is as follows:
\begin{itemize}
    \item \textit{Filling stage}: We use all data with a sampling rate of at least \SI{16}{kHz}, with all samples downsampled to 16 kHz for training. The input consists of noisy speech degraded by packet loss distortion, with other distortion types introduced at a predefined probability. The target output is the corresponding sample with packet loss distortion removed.

    \item \textit{Separation stage}: Similar to the filling stage, we use all data with a sampling rate of no less than \SI{16}{kHz}, with all samples downsampled to \SI{16}{kHz} for training. The input is speech impaired by noise, reverberation, clipping, and wind noise distortion, with additional distortion types introduced at a specific probability (excluding packet loss distortion). The target output is the corresponding sample free from noise, reverberation, and clipping distortion.

    \item \textit{Restoration stage}: We exclusively use data with a sampling rate of \SI{48}{kHz}. The input is speech corrupted by bandwidth limitation, codec artifacts, and packet loss distortion. The target output is the corresponding clean sample.

\end{itemize}

The official validation set comprises 1,000 noisy-clean data pairs, simulated in the same manner as the training data. The official blind set, used for the final ranking, comprises 900 noisy utterances, with 50\% being real recordings and the remaining synthetic. Additionally, it includes data in an unseen language, data from unseen corpora, and data distorted by previously unencountered distortion types.

\begin{table}[tbp]
\centering
\caption{Configuration of TF-GridNet-S hyperparameters}
\vspace{-5pt}
\resizebox{\linewidth}{!}{
\begin{tabular}{ccc}
%\hline
\toprule
\textbf{Symbols} & \textbf{Descriptions}     & \textbf{Value} \\ \midrule
$D$     & Embedding dimension                         & 48    \\
$B$     & Number of blocks                            & 5     \\ \midrule
$I$     & Kernel size for Unfold                      & 4     \\
$J$     & Stride size for Unfold                      & 1     \\
$H$     & Number of hidden units in LSTM              & 100   \\ \midrule
$E$     & Number of output channels in self-attention & 2     \\
$L$     & Number of heads in self-attention           & 4     \\
\bottomrule
\end{tabular}
}
\label{tab:1}
\end{table}

\begin{table}[tbp]
    \centering
    \caption{Training configurations of different stages.}
    \vspace{-5pt}
    \resizebox{\linewidth}{!}{
	\begin{tabular}{ccccccc}
		\toprule
		\textbf{Stage}       & \textbf{Bs} & \textbf{Len}  & \textbf{Ts} & \textbf{Ws} & \textbf{Minlr} & \textbf{Maxlr} \\ \midrule
	  Fill     & 6  & 2 s   & 50,000  & 5,000  & 1e-6    & 1e-3    \\ \midrule
	Sep  & 1   & 4 s   & 200,000  & 20,000  & 1e-6    & 1e-3    \\
	MAFT  & 1   & 4 s   & 5,000   & 1,500    & 1e-6    & 1e-4    \\ \midrule
	Res &  6  & 2 s    & 100,000  & 10,000  & 1e-6    & 5e-4    \\
	SAFT/JFT &  6  & 2 s    & 25,000   & 2,500    & 1e-6    & 1e-4    \\ \bottomrule
	\end{tabular}
     }
\label{tab:2}
\end{table}

\subsection{Experimental setup}
The STFT is computed using a Hann window with a length of 32 ms, a hop length of 16 ms, and an FFT size equal to the window size. The hyperparameter configuration of TF-GridNet-S employed in the filling stage is detailed in Table~\ref{tab:1}. For TF-GridNet-L utilized in the separation stage, the parameters \(I\) and \(J\) remain the same as those in TF-GridNet-S, while all other hyperparameters are doubled in size. The TF-GridNet module in CWS-TF-GridNet in the restoration stage adopts the same configuration as TF-GridNet-S. 

The models in each stage are trained independently on 8 NVIDIA 4090 GPUs. All models use the AdamW optimizer with a linear-warm-up-cosine-annealing scheduler for learning rate adjustment. Table \ref{tab:2} presents a detailed overview of the training configurations across different stages (Fill for filling, Sep for separation, and Res for restoration), including the MAFT and SAFT/JFT phases. These configurations include batch size per GPU (Bs), utterance length (Len), training steps (Ts), warm-up steps (Ws), and the specified minimum (Minlr) and maximum learning rates (Maxlr). The whole framework has \SI{30.40}{M} parameters, with a computational cost of \SI{493.0}{GMACs} per second for 16-kHz input, and \SI{1360.5}{GMACs} per second for 48-kHz input.

\begin{table*}[tbp]
\centering
\caption{Results of the ablation study on the official validation set.} 
\vspace{-5pt}
\label{tab:3}
\renewcommand{\arraystretch}{1.15}
\resizebox{\linewidth}{!}{
\begin{tabular}{c|c|ccc|ccccc|cc|cc}
\toprule
  \textbf{IDs} &
  \textbf{Methods} &
  \textbf{DNSMOS} $\uparrow$ &
  \textbf{NISQA} $\uparrow$ &
  \textbf{UTMOS} $\uparrow$ &
  \textbf{PESQ} $\uparrow$ &
  \textbf{ESTOI} \cite{ESTOI} $\uparrow$ &
  \textbf{SDR (dB)} $\uparrow$ &
  \textbf{MCD} $\downarrow$ &
  \textbf{LSD} $\downarrow$ &
  \textbf{SpeechBERTScore} $\uparrow$ &
  \textbf{PhnSim} $\uparrow$ &
  \textbf{SpkSim} $\uparrow$ &
  \textbf{CAcc (\%)} $\uparrow$ \\ \midrule
-  & Noisy                          & 1.78 & 1.56 & 1.51 & 1.26 & 0.57 & 3.28 & 8.94 & 5.91 & 0.72 & 0.60 & 0.59 & 79.49 \\ \hline
1  & Sep+Res                        & 3.05 & 3.48 & 2.47 & 2.49 & 0.80 & 13.03 & 3.86 & 3.20 & 0.82 & 0.82 & 0.79 & 87.32 \\
2  & Fill+Sep+Res                   & 3.08 & 3.67 & 2.54 & 2.54 & 0.81 & 13.19 & 3.73 & 3.01 & 0.83 & 0.83 & 0.83 & 86.23 \\ \hline
3  & Fill+Sep                       & 3.08 & 3.59 & 2.52 & 2.56 & 0.81 & 13.28 & 4.03 & 3.88 & 0.83 & 0.83 & 0.76 & 86.53\\
4  & Fill+Sep-MAFTv1                  & 3.07 & 3.44 & 2.52 & 2.54 & 0.81 & 13.27 & 3.62 & 3.81 & 0.83 & 0.83 & 0.77 & 86.37\\
5  & Fill+Sep-MAFTv2                  & 3.07 & 3.47 & 2.51 & 2.60 & 0.81 & 13.27 & 3.65 & 3.84 & 0.83 & 0.83 & 0.77 & 86.49\\
6  & Fill+Sep-MAFTv3                  & 3.14 & 3.69 & 2.71 & 2.59 & 0.81 & 13.23 & 3.69 & 3.86 & 0.83 & 0.83 & 0.77 & 85.74\\
7  & Fill+Sep-MAFTv4                  & 3.13 & 3.64 & 2.70 & 2.58 & 0.81 & 13.23 & 3.69 & 3.83 & 0.83 & 0.82 & 0.77 & 85.72\\
8  & Filll+Sep-MAFTv5                 & 3.13 & 3.66 & 2.70 & 2.59 & 0.81 & 13.23 & 3.68 & 3.82 & 0.83 & 0.82 & 0.77 & 85.82\\ \hline
9  & Fill+Sep-MAFTv3+Res              & 3.14 & 3.77 & 2.72 & 2.58 & 0.81 & 13.15 & 3.50 & 2.85 & 0.83 & 0.82 & 0.80 & 85.27 \\
10 & Fill+Sep-MAFTv3+Res-SAFT         & 3.14 & 3.88 & 2.70 & 2.39 & 0.80 & 12.16 & 3.75 & 2.93 & 0.82 & 0.81 & 0.79 & 84.19 \\ 
11 & Fill+Sep-MATFv3+Res-JFT          & 3.14 & 3.92 & 2.76 & 2.48 & 0.80 & 11.95 & 3.77 & 2.97 & 0.83 & 0.82 & 0.80 & 84.36 \\ \hline
12 & Fill+Sep-MATFv3+Res-JFT+BWAI        & 3.14 & 3.86 & 2.75 & 2.56 & 0.80 & 12.71 & 3.60 & 2.90 & 0.83 & 0.82 & 0.80 & 84.69 \\ 
\bottomrule
\end{tabular}
}
\end{table*}

\begin{table*}[tbp]
\centering
\caption{Results of the evaluations on the official blind test set. 
\vspace{-5pt}
\textbf{BOLD} indicates the best score in each metric.} 
\label{tab:4}
\renewcommand{\arraystretch}{1.15}
\resizebox{\linewidth}{!}{
\begin{tabular}{c|ccc|cccccc|cc|cc|c}
\toprule
  \textbf{Methods} &
  \textbf{DNSMOS} $\uparrow$ &
  \textbf{NISQA} $\uparrow$ &
  \textbf{UTMOS} $\uparrow$ &
  \textbf{POLQA} \cite{POLQA} $\uparrow$ &
  \textbf{PESQ} $\uparrow$ &
  \textbf{ESTOI} $\uparrow$ &
  \textbf{SDR (dB)} $\uparrow$ &
  \textbf{MCD} $\downarrow$ &
  \textbf{LSD} $\downarrow$ &
  \textbf{SpeechBERTScore} $\uparrow$ &
  \textbf{PhnSim} $\uparrow$ &
  \textbf{SpkSim} $\uparrow$ &
  \textbf{CAcc (\%)} $\uparrow$ &
  \textbf{MOS} $\uparrow$ \\ \midrule
Noisy                        & 1.90 & 1.58 & 1.55 & 1.83 & 1.31 & 0.58 & 3.24 & 9.34 & 5.84 & 0.70 & 0.51 & 0.55 & 73.41 & 2.13 \\
Baseline                     & 2.85 & 2.77 & 1.92 & 2.99 & 2.24 & 0.76 & 10.24 & \textbf{3.80} & \textbf{2.72} & 0.82 & 0.67 & 0.70 & 75.60 & 2.96 \\ \hline
TS-URGENet                   & \textbf{2.92} & \textbf{3.24} & \textbf{2.16} & \textbf{3.17} & \textbf{2.47} & \textbf{0.79} & \textbf{11.10} & 3.96 & 2.99 & \textbf{0.84} & \textbf{0.70} & \textbf{0.74} & \textbf{76.06} & \textbf{3.32} \\
\bottomrule
\end{tabular}
}
\end{table*}

\section{Results and Analysis}
In this section, we evaluate the proposed system using the officially specified metrics\footnote{\url{https://urgent-challenge.github.io/urgent2025/rules/}}.

\subsection{Ablation study}
\subsubsection{Effects of the filling stage}
A key novelty of our proposed three-stage framework is the positioning of the filling stage first, which is designed to fill the packet loss regions under noise interference. To validate the effectiveness of this approach, we compare two scenarios: separation-restoration and filling-separation-restoration. The results are presented in Table \ref{tab:3} (ID1 and ID2). It is evident that performing the filling stage prior to the separation and restoration stages improves all the metrics except for CAcc, which shows a minor decline. Notably, significant enhancements are observed in NISQA, UTMOS, PESQ, MCD, and LSD.

\subsubsection{Effects of metric-aware fine-tuning}
\label{sec:maft}
While metric-specific loss functions effectively optimize individual metrics, multi-metric scenarios often lead to conflicting gradient directions, requiring careful analysis. We investigate five progressively enhanced metric-aware loss configurations:
\begin{itemize}
    \item MAFTv1: add MCD-aware term;
    \item MAFTv2: add MCD and PESQ-aware terms;
    \item MAFTv3: add MCD, PESQ, and UTMOS-aware terms;
    \item MAFTv4: add MCD, PESQ, UTMOS, and DNSMOS-aware terms;
    \item MAFTv5: add MCD, PESQ, UTMOS, DNSMOS, and WavLM-aware terms.
\end{itemize}
The results are presented in Table \ref{tab:3} (ID3-8). A comparison between ID3 and ID4 reveals that the MCD-aware component significantly enhances MCD and slightly improves LSD. The addition of the PESQ-aware term (ID5) boosts PESQ and contributes to a minor recovery in NISQA, albeit at the cost of a slight degradation in MCD and LSD. Incorporating the UTMOS-aware term (ID6) leads to notable improvements in all non-intrusive metrics but causes a slight decline in other metrics, including SDR, MCD, LSD, and CAcc. Further inclusion of the DNSMOS-aware term (ID7) and WavLM-aware term (ID8) does not yield improvements in the corresponding metrics. These findings suggest that different metric-aware loss components may conflict with each other. Nevertheless, by appropriately combining these components, substantial improvements can still be achieved. 

\subsubsection{Effects of joint stage and metric-aware fine-tuning}
A comparison between ID6 and ID9 highlights the effectiveness of the restoration stage, which enhances nearly all metrics, with a particularly significant improvement in LSD. Further fine-tuning using the SAFT strategy (ID10) results in an additional boost in NISQA, but causes a noticeable decline in several other metrics, especially PESQ. However, it is worth noting that SAFT delivers a clear improvement in subjective listening quality, which cannot be fully captured in the table. By replacing SAFT with our proposed JFT (ID11), NISQA and UTMOS show further gains, and PESQ obtains a significant recovery. Nevertheless, SNR, MCD, and LSD exhibit a decline once again.

Since we have observed that the subjective quality improvement brought by fine-tuning is more evident for bandwidth-limited speech, we employ a bandwidth-aware inference (BWAI) strategy to balance objective metrics and subjective quality. During inference, the speech identified as bandwidth-limited undergoes the fine-tuned restoration module; otherwise, it is processed through the original restoration module. The result, as shown in ID12, demonstrates an improvement in NISQA and UTMOS compared to ID9, along with an enhancement in subjective quality, although at the cost of hampering some intrusive metrics, including SDR, MCD, and LSD. Considering the trade-off between objective metrics and subjective quality, we ultimately selected ID12 as our final submission.

\subsection{Evaluation on the official blind test set}
Evaluation results on the official blind test set are presented in Table \ref{tab:4}. Compared to the baseline, TS-URGENet achieves significant improvements in terms of all metrics except for MCD and LSD, which shows the superiority of our proposed system.

\section{Conclusion}
This paper presents our submission to the Interspeech 2025 URGENT Challenge. We introduce a three-stage framework capable of handling input speech with different distortions and sampling frequencies. Additionally, we propose an effective metric-aware fine-tuning function to boost various evaluation metrics specified in the challenge. Our system achieves outstanding performance in the Interspeech 2025 URGENT Challenge, ranking 2nd in Track 1.

\section{Acknowledgment}
This work was supported by the National Natural Science Foundation of China (Grant No. 12274221) and the Yangtze River Delta Science and Technology Innovation Community Joint Research Project (Grant No. 2024CSJGG1100).

\bibliographystyle{IEEEtran}
\bibliography{ref.bib}

\end{document}